\documentclass[aps,preprint,showpacs,preprintnumbers,amsmath,amssymb]{revtex4}
\usepackage{amsmath,mathrsfs,amsbsy,color,graphicx,bm,amsthm,amsfonts}
\usepackage{units}
\usepackage{bbm}
\usepackage{times}
\usepackage{dcolumn}
\usepackage{mathrsfs}
\usepackage{amsmath,amssymb,epsfig}
\begin{document}

\title{ Non-monotonic evolution of multipartite entanglement under the Unruh effect }
\author{ Shu-Min Wu$^1$\footnote{Email: smwu@lnnu.edu.cn}, Si-Han Shang$^1$, Si-Yu Liu$^1$, Rui-Yang Xu$^1$, Qianqian Liu\footnote{Email:  qianqianliu@xtu.edu.cn  (corresponding author)}$^2$,  Xiao-Li Huang$^1$\footnote{Email: huangxiaoli1982@foxmail.com (corresponding author)} }
\affiliation{$^1$  Department of Physics, Liaoning Normal University, Dalian 116029, China \\
$^2$ Department of Physics, Xiangtan University, Xiangtan 411105, China
}


\begin{abstract}
We investigate the behavior of tetrapartite entanglement in a four-qubit Dicke state under relativistic motion by employing the Unruh-DeWitt detector model, where one detector undergoes uniform acceleration. We show that the entanglement exhibits a non-monotonic evolution: it first decreases and subsequently increases toward a finite value as the acceleration grows. In contrast to the conventional view that the Unruh effect leads to a monotonic degradation of multipartite entanglement, our results demonstrate that it can instead enhance multipartite entanglement within a finite parameter regime. This behavior reveals a dual role of the Unruh effect in multipartite systems. Our findings therefore provide a refined understanding of relativistic multipartite quantum correlations, indicating that Dicke states constitute more robust multipartite quantum resources against Unruh-induced decoherence and may offer advantages for relativistic quantum information processing tasks.
\end{abstract}

\vspace*{0.5cm}
\maketitle
\section{Introduction}
Quantum entanglement is a central resource in quantum information science  \cite{Q1}, enabling tasks such as quantum teleportation \cite{Q12,Q13,Q14}, quantum computation \cite{Q15,Q16}, quantum cryptography \cite{Q17}, and quantum communication \cite{Q19,Q20,Q21}. Among multipartite entangled states, Dicke states play a distinguished role due to their unique structural and operational properties. Originally introduced to describe collective light emission from atomic ensembles  \cite{Q22}, Dicke states exhibit remarkable robustness against particle loss \cite{Q23} and maintain a high degree of entanglement persistence compared to paradigmatic states such as GHZ and W states. In addition, their permutational symmetry significantly simplifies quantum state tomography \cite{Q24,Q25} and entanglement characterization \cite{Q26,Q27,Q28,Q29}. From a resource-theoretic perspective, Dicke states are highly versatile: they can be converted into other classes of multipartite entangled states, including GHZ and W states, via local operations and classical communication \cite{Q30,Q31}, and have been shown to support a variety of quantum information processing tasks such as multiparty quantum networking protocols \cite{Q32}. These features make Dicke states particularly attractive candidates for exploring the behavior of multipartite entanglement in relativistic settings.

Relativistic quantum information \cite{ZZ23,ZZ24,ZZ25,ZZ26,ZZ27,ZZ28,ZZ29,ZZ30,ZZ31,QTO1,QTO2,QTO3,QTO4,QTO5,ADD1,ADD2,ADD3,ADD4,ADD5,ADD6,ADD7,ADD8,ADD9,ADD10} is an interdisciplinary framework that combines the spacetime geometry and gravitational structure described by general relativity, the observer-dependent quantum dynamics formulated in quantum field theory in curved spacetime, and the resource-based perspective of quantum information theory, with the aim of systematically investigating how relativistic effects influence quantum entanglement and other quantum correlations relevant to quantum information processing \cite{SDF1,SDF2,SDF3,SDF4,SDF5,SDF6,SDF7,SDF8,SDF9,SDF10,SDF11,SDF12,SDF13,SDF14,SDF15,SDF16,SDF17,SDF18,SDF19,SDF20,SDF21,SDF22,SDF23,SDF24,SDF25,SDF26,SDF29,SDF30,ZZ1,ZZ2,ZZ3,ZZ6,ZZ7,ZZ8,ZZ9,ZZ10,ZZ11,ZZ12,ZZ13,ZZ14,ZZ15,ZZ16,ZZ17,ZZ18,ZZ19,ZZ20,ZZ21}. Previous studies have shown that, within free quantum field models, multipartite entanglement,  steering, and  coherence exhibit a monotonic degradation with increasing acceleration and Hawking temperature parameters \cite{SDF27,SDF28,SDF31,ZZ4,ZZ5,RM1,RM2,RM3,RM4,RM5,RM6}. However, the free-field model is inherently an idealized theoretical description in which field modes are globally excited throughout the entire spacetime. Such nonlocal excitation schemes are difficult to realize in practical experimental settings, thereby limiting the experimental testability and physical feasibility of these conclusions. Therefore, we introduce the Unruh-DeWitt detector as our theoretical framework. The  Unruh-DeWitt detector describes the interaction between a finite-dimensional quantum system and a quantum field through a local coupling \cite{UD20,UD2,UD3,UD4,UD5,UD6}, thereby avoiding the idealized assumption of global field-mode excitation inherent in free-field models. This feature makes it a physically realistic and operationally well-defined tool for investigating the evolution of quantum resources in non-inertial motion and curved spacetime backgrounds. Previous studies within this model have shown that multipartite entanglement and coherence gradually degrade with increasing acceleration parameters, and may even undergo entanglement sudden death in certain parameter regimes  \cite{UD20,UD21}.

Although Dicke states play an important role in quantum information theory, their properties in relativistic frameworks have not yet been systematically and thoroughly investigated. Previous studies have generally suggested that relativistic effects tend to degrade global multipartite entanglement and may even lead to irreversible loss of quantum correlations, giving rise to the widely accepted view that multipartite quantum resources typically exhibit monotonic decay under relativistic effects \cite{SDF27,SDF28,SDF31,ZZ4,ZZ5,RM1,RM2,RM3,RM4,RM5,RM6,UD20}. However, owing to the distinctive structure of entanglement distribution in Dicke states, it remains unclear whether their multipartite entanglement under relativistic conditions still strictly follows this monotonic degradation behavior. This naturally leads to our first motivation, namely, to examine whether the multipartite entanglement of Dicke states can deviate from the conventional monotonic decay picture. Furthermore, given their known robustness against particle loss and favorable entanglement structure, our second motivation is to explore whether Dicke states can serve as more robust multipartite quantum resources against Unruh-induced decoherence. Addressing these questions is therefore not only important for testing the generality of the conventional understanding, but also for identifying advantageous quantum resources for relativistic quantum information processing.

In this paper, we investigate the dynamics of tetrapartite entanglement in a four-qubit Dicke state within a relativistic framework, where one of the Unruh-DeWitt detectors (David) undergoes uniform acceleration. Initially, Alice, Bob, Charlie, and David share a Dicke state in flat Minkowski spacetime, and the Unruh effect induced by David’s acceleration is incorporated to analyze its impact on quantum correlations. We focus on the evolution of the global multipartite entanglement as functions of the acceleration parameter. Our results show that the entanglement dynamics do not follow the commonly assumed monotonic decay behavior. Instead, the multipartite entanglement of the Dicke state exhibits a non-monotonic evolution: it first decreases and subsequently increases toward a finite value as the acceleration grows. This behavior demonstrates that, under relativistic effects, multipartite entanglement can not only degrade but also be enhanced within a certain parameter regime. Furthermore, in contrast to GHZ and W states, whose multipartite entanglement typically suffers from sudden death under relativistic effects \cite{UD20}, the Dicke state retains a nonvanishing amount of entanglement even in the large-acceleration limit. This highlights the superior robustness of Dicke-state entanglement and indicates that it constitutes a more resilient multipartite quantum resource for relativistic quantum information processing.

This paper is organized as follows. In Sec.~II, we present the evolution of a four-qubit Dicke state in which one detector undergoes uniform acceleration. Sec.~III analyzes the dynamics of the  $1-3$   entanglement and the global quantum entanglement under relativistic motion. Finally, Sec.~IV summarizes our main results and draws the conclusions.

\section{The evolution of a tetrapartite system with an accelerated detector }
\begin{figure}
\centering
\includegraphics[height=2.1in,width=5.6in]{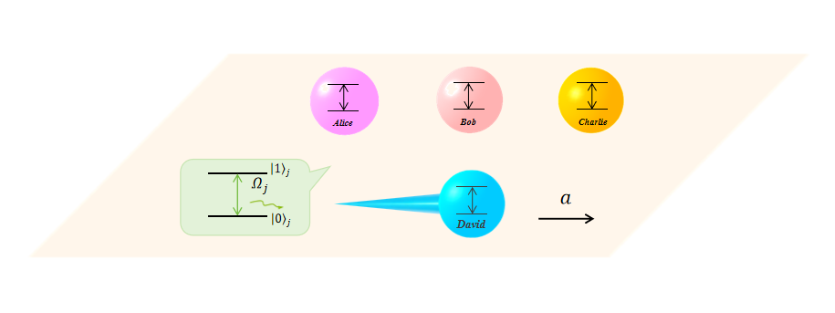}
\caption{Schematic setup of the tetrapartite Unruh--DeWitt detector system. Four observers (Alice, Bob, Charlie, and David) share an initial four-qubit Dicke state, while only David undergoes uniform acceleration with proper acceleration $a$, and the remaining detectors remain inertial. Each detector is modeled as a two-level system with energy gap $\Omega$.}
\label{F1}
\end{figure}

In this section, we analyze the dynamics of a tetrapartite Unruh–DeWitt detector system under relativistic motion. The initial four-qubit Dicke state $D_4$ is given by
\begin{eqnarray}\label{w1}
|\Psi_{ABCD}\rangle=\frac{1}{\sqrt{6}}(|0011\rangle+|0101\rangle+|1001\rangle+
|1100\rangle+|0110\rangle+|1010\rangle).
\end{eqnarray}
The system consists of four observers, Alice, Bob, Charlie, and David, each equipped with a Unruh-DeWitt  detector. We assume that only David undergoes uniform acceleration $a$ for a finite proper time interval $\Delta$, while the remaining detectors stay inertial and are switched off, as shown in Fig.\ref{F1}. During this interval, David's detector is switched on and interacts locally with a scalar field. Its worldline is parametrized by
\begin{eqnarray}\label{w2}
t(\tau)=a^{-1}\sinh a\tau,  \quad x(\tau)=a^{-1}\cosh a\tau,  \quad  y(\tau)=z(\tau)=0,
\end{eqnarray}
where $\tau$ denotes the proper time \cite{ZZ21}. Throughout this work, we set $c=\hbar=k_{B}=1$.
The initial state of the detector-field system is taken as
\begin{eqnarray}\label{w3}
|\Psi_{-\infty}^{ABCD}\rangle=|\Psi_{ABCD}\rangle\otimes|0_{M}\rangle,
\end{eqnarray}
where $|0_{M}\rangle$ is the Minkowski vacuum of the external scalar field. The total Hamiltonian reads
\begin{eqnarray}\label{w5}
H_{ABCD\phi}=H_{A}+H_{B}+H_{C}+H_{D}+H_{KG}+H_{\textrm{int}}^{D\phi},
\end{eqnarray}
where $H_{KG}$ denotes the free massless scalar field Hamiltonian, and the internal Hamiltonian of each detector is given by $H_S=\Omega S^\dagger S$ ($S=A,B,C,D$), with $\Omega$ denoting the energy gap between the ground state $|0\rangle$ and excited state $|1\rangle$ \cite{UD21}. The ladder operators satisfy $S^\dagger|0\rangle=|1\rangle$, $S|1\rangle=|0\rangle$, and $S|0\rangle=S^\dagger|1\rangle=0$. The  interaction Hamiltonian between David's detector and the scalar field is described by
\begin{eqnarray}\label{w4}
H_{\textrm{int}}^{D\phi}(t)=\epsilon(t)\int_{\sum_{t}}\textrm{d}^{3} \boldsymbol{x}\sqrt{-g}\phi(x)[\chi(\boldsymbol{x})D+\bar{\chi}(\boldsymbol{x})D^{\rm \dagger}],
\end{eqnarray}
where $g \equiv \det(g_{ab})$ with $g_{ab}$ the Minkowski metric. The smearing function is chosen as a Gaussian profile,
$\chi(\boldsymbol{x}) = (\kappa\sqrt{2\pi})^{-3}\textrm{exp}(\boldsymbol{-x}^{2}/2\kappa^{2})$,
which localizes the interaction within a finite spatial region around the detector.

Within the weak-coupling regime, the final state of the detector-field system can be obtained perturbatively to first order in the coupling strength $\epsilon$. The time-evolved state is given by
\begin{eqnarray}\label{w6}
|\Psi^{ABCD\phi}_\infty\rangle=\{I-i[\phi(f)D+\phi(f)^{\dag}D^{\dag}]\}|\Psi_{-\infty}^{ABCD\phi}\rangle,
\end{eqnarray}
where the smeared field operator is defined as $\phi(f) \equiv \int d^{4}x\sqrt{-g}\chi(x)f$, with the smearing function $f\equiv\epsilon(t)\textrm{e}^{-\textrm{i}\Omega t}\chi(\boldsymbol{x})$  being compactly supported in Minkowski spacetime  \cite{UD6}. Equivalently, $\phi(f)$ can be expressed in terms of Rindler modes as $\phi(f)=i\left[a_{RI}(\overline{uE\overline{f}})-a^{\dagger}_{RI}(uEf)\right]$, where $a_{RI}$ and $a^{\dagger}_{RI}$ denote the annihilation and creation operators associated with Rindler modes, respectively, $u$ projects onto the positive-frequency sector of the Klein--Gordon solutions in Rindler spacetime, and $E$ is the Pauli--Jordan function, defined as the difference between the advanced and retarded Green's functions.

By substituting the initial state in Eq.(\ref{w3}) into Eq.(\ref{w6}), the final state of the total system can be expressed in terms of Rindler operators as
\begin{eqnarray}\label{w9}
|\Psi^{ABCD\phi}_{\infty}\rangle&=&|\Psi^{ABCD\phi}_{-\infty}\rangle+\frac{1}{\sqrt{6}}
[(|0010\rangle+|0100\rangle+|1000\rangle)\otimes(a^{\dagger}_{RI}(\lambda)|0_{M}\rangle)\\
&+&(|1101\rangle+|0111\rangle+|1011\rangle)\otimes(a_{RI}
(\overline{\lambda})|0_{M}\rangle)]\notag,
\end{eqnarray}
where $\lambda=-uEf$, and $a_{RI}(\overline\lambda)$, $a^{\dagger}_{RI}({\lambda})$ denote annihilation and creation operators defined in Rindler region $\emph{I}$, while $|0_{M}\rangle$ is the Minkowski vacuum. The Bogoliubov transformations relating Rindler and Minkowski operators take the form
\begin{eqnarray}\label{w10}
a_{RI}(\overline{\lambda})=\frac{a_{M}(\overline{F_{1\Omega}})+\textrm{e}^{-\pi\Omega/a}
a^{\dagger}_{M}(F_{2\Omega})}{(1-\textrm{e}^{-2\pi\Omega/a})^{1/2}},
\end{eqnarray}
\begin{eqnarray}\label{w11}
a^{\dagger}_{RI}(\lambda)=\frac{a^{\dagger}_{M}({F_{1\Omega}})+
\textrm{e}^{-\pi\Omega/a}a_{M}(\overline{F_{2\Omega}})}{(1-\textrm{e}^{-2\pi\Omega/a})^{1/2}},
\end{eqnarray}
where $F_{1\Omega} = \frac{\lambda + e^{-\pi \Omega / a} \lambda \circ \omega}{\sqrt{1 - e^{-2\pi \Omega / a}}}$ and $F_{2\Omega} = \frac{\overline{\lambda \circ \omega} + e^{-\pi \Omega / a}\,\overline{\lambda}}{\sqrt{1 - e^{-2\pi \Omega / a}}}$, with  $\omega(t,x,y,z) = (-t, -x, y, z)$ denoting the wedge reflection that maps region $I$ to region $II$, and $\circ$ representing composition.

Using the Bogoliubov transformations in Eqs.(\ref{w10}) and (\ref{w11}), the final state in Eq.(\ref{w9}) can be rewritten as
\begin{eqnarray}\label{w16}
|\Psi^{ABCD\phi}_{\infty}\rangle&=&|\Psi^{ABCD\phi}_{-\infty}\rangle+\frac{1}
{\sqrt{6}}\nu\bigg[\frac{(|0010\rangle+|0100\rangle+|1000\rangle)\otimes|1_{\tilde{F}_{1\Omega}}\rangle}
{(1-\textrm{e}^{-2\pi\Omega/a})^{1/2}}\\
&+&\textrm{e}^{-\pi\Omega/a}\frac{(|1101\rangle+|0111\rangle+|1011\rangle)\otimes
|1_{\tilde{F}_{2\Omega}}\rangle}{(1-\textrm{e}^{-2\pi\Omega/a})^{1/2}}\bigg],
\end{eqnarray}
where $\tilde{F}_{\textrm{i}\Omega}=F_{\textrm{i}\Omega}/\nu$.
Tracing over the scalar field degrees of freedom yields the reduced density matrix of the detectors,
\begin{eqnarray}\label{w18}
\rho_{ABCD}=\|\Psi^{ABCD\phi}_{\infty}\|^{-2}\textrm{tr}_{\phi}|\Psi^{ABCD\phi}_{\infty}\rangle
\langle\Psi^{ABCD\phi}_{\infty}|,
\end{eqnarray}
where $\|\Psi^{ABCD\phi}_{\infty}\|^{2}=1+\frac{\upsilon^{2}(1+e^{-2\pi\Omega/a})}{2(1-e^{-2\pi\Omega/a})}$ guarantees that the final density matrix is properly normalized, i.e., $\textrm{tr}\rho_{ABCD}=1$.
With the acceleration parameter defined as $q\equiv e^{-2\pi\Omega/a}$, which is a monotonically increasing function of the acceleration $a$, the limits $q\rightarrow0$ and $q\rightarrow1$ correspond to vanishing and infinite acceleration, respectively. In the subspace spanned by \{$|0010\rangle$, $|0011\rangle$, $|0100\rangle$, $|0101\rangle$, $|0110\rangle$, $|0111\rangle$, $|1000\rangle$, $|1001\rangle$, $|1010\rangle$, $|1011\rangle$, $|1100\rangle$, and $|1101\rangle$\}, the final reduced density matrix of the detectors can be written in block-matrix form as
\begin{eqnarray}\label{w20}
\scalebox{0.8}{$
\rho_{ABCD}=
\left(\!\!\begin{array}{cccccccccccccccc}
Z_{1}&0&Z_{1}&0&0&0&Z_{1}&0&0&0&0&0\\
0&Z_{0}&0&Z_{0}&Z_{0}&0&0&Z_{0}&Z_{0}&0&Z_{0}&0\\
Z_{1}&0&Z_{1}&0&0&0&Z_{1}&0&0&0&0&0\\
0&Z_{0}&0&Z_{0}&Z_{0}&0&0&Z_{0}&Z_{0}&0&Z_{0}&0\\
0&Z_{0}&0&Z_{0}&Z_{0}&0&0&Z_{0}&Z_{0}&0&Z_{0}&0\\
0&0&0&0&0&Z_{2}&0&0&0&Z_{2}&0&Z_{2}\\
Z_{1}&0&Z_{1}&0&0&0&Z_{1}&0&0&0&0&0 \\
0&Z_{0}&0&Z_{0}&Z_{0}&0&0&Z_{0}&Z_{0}&0&Z_{0}&0\\
0&Z_{0}&0&Z_{0}&Z_{0}&0&0&Z_{0}&Z_{0}&0&Z_{0}&0\\
0&0&0&0&0&Z_{2}&0&0&0&Z_{2}&0&Z_{2}\\
0&Z_{0}&0&Z_{0}&Z_{0}&0&0&Z_{0}&Z_{0}&0&Z_{0}&0\\
0&0&0&0&0&Z_{2}&0&0&0&Z_{2}&0&Z_{2}\\
\end{array}\!\!\right)
$},
\end{eqnarray}
where the parameters $Z_{0}$, $Z_{1}$, and $Z_{2}$ are given by
\begin{eqnarray}\label{w21}
Z_{0}=\frac{1-q}{3\nu^{2}(1+q)+6(1-q)}\notag,
\end{eqnarray}
\begin{eqnarray}\label{w22}
Z_{1}=\frac{\nu^{2}}{3\nu^{2}(1+q)+6(1-q)}\notag,
\end{eqnarray}
\begin{eqnarray}\label{w23}
Z_{2}=\frac{\nu^{2}q}{3\nu^{2}(1+q)+6(1-q)}\notag,
\end{eqnarray}
respectively.
The effective coupling is defined as $\nu^{2} \equiv \|\lambda\|^{2} = \frac{\epsilon^{2}\Omega\Delta}{2\pi}\, e^{-\Omega^{2}\kappa^{2}}$, subject to the condition $\Omega^{-1} \ll \Delta$, which is required for the validity of this definition. Moreover, for the perturbative treatment adopted in this work to remain valid, the effective coupling must satisfy $\nu^{2} \ll 1$.

\section{Tetrapartite  entanglement of the four-qubit Dicke state in relativistic motion}
\subsection{ $1-3$  bipartite entanglement in relativistic motion}
To quantify entanglement, we employ the negativity, which is a widely used measure based on the positive partial transpose criterion. A quantum state $\rho$ is entangled if the partial transpose of its density matrix possesses at least one negative eigenvalue. For the tetrapartite system, we consider the negativities defined as
\begin{eqnarray}\label{ww26}
N_{A(BCD)}=\parallel\rho^{T_{A}}_{A(BCD)}\parallel-1,
N_{AB}=\parallel\rho^{T_{A}}_{AB}\parallel-1,
\end{eqnarray}
which characterize the  $1-3$ and $1-1$ bipartite entanglement ( $1-3$ tangle and $1-1$ tangle), respectively, where $\rho^{T_{A}}$ denotes the partial transpose with respect to subsystem $A$, and $\|\cdot\|$ is the trace norm \cite{NE1,NE2}. For a Hermitian operator $M$, the trace norm can be expressed as $\|M\|=\mathrm{tr}\sqrt{M^\dagger M}$, and equivalently the negativity can be written as
\begin{eqnarray}\label{ww266}
\|M\|-1=2\sum_{i}|\lambda^{(-)}_{M,i}|,
\end{eqnarray}
where $\lambda^{(-)}_{M,i}$ denote the negative eigenvalues of $M$.

To evaluate the negativities, we perform the partial transpose of the reduced density matrix with respect to subsystem $A$, yielding
\begin{eqnarray}\label{w25}
\scalebox{0.7}{$
\rho^{T_{A}}_{ABCD}=
\left(\!\!\begin{array}{cccccccccccccccc}
0&0&0&0&0&0&0&0&0&0&Z_{1}&0&Z_{1}&0&0&0\\
0&0&0&0&0&0&0&0&0&0&0&Z_{0}&0&Z_{0}&Z_{0}&0\\
0&0&Z_{1}&0&Z_{1}&0&0&0&0&0&0&Z_{0}&0&Z_{0}&Z_{0}&0\\
0&0&0&Z_{0}&0&Z_{0}&Z_{0}&0&0&0&0&0&0&0&0&Z_{2}\\
0&0&Z_{1}&0&Z_{1}&0&0&0&0&0&0&Z_{0}&0&Z_{0}&Z_{0}&0\\
0&0&0&Z_{0}&0&Z_{0}&Z_{0}&0&0&0&0&0&0&0&0&Z_{2}\\
0&0&0&Z_{0}&0&Z_{0}&Z_{0}&0&0&0&0&0&0&0&0&0\\
0&0&0&0&0&0&0&Z_{2}&0&0&0&0&0&0&0&0\\
0&0&0&0&0&0&0&0&Z_{1}&0&0&0&0&0&0&0\\
0&0&0&0&0&0&0&0&0&Z_{0}&Z_{0}&0&Z_{0}&0&0&0\\
Z_{1}&0&0&0&0&0&0&0&0&Z_{0}&Z_{0}&0&Z_{0}&0&0&0\\
0&Z_{0}&Z_{0}&0&Z_{0}&0&0&0&0&0&0&Z_{2}&0&Z_{2}&0&0\\
Z_{1}&0&0&0&0&0&0&0&0&Z_{0}&Z_{0}&0&Z_{0}&0&0&0\\
0&Z_{0}&Z_{0}&0&Z_{0}&0&0&0&0&0&0&Z_{2}&0&Z_{2}&0&0\\
0&Z_{0}&Z_{0}&0&Z_{0}&0&0&0&0&0&0&0&0&0&0&0\\
0&0&0&Z_{2}&0&Z_{2}&0&0&0&0&0&0&0&0&0&0\\
\end{array}\!\!\right)
$}.
\end{eqnarray}
Similarly, in the subspace spanned by \{$|0010\rangle$, $|0011\rangle$, $|0100\rangle$, $|0101\rangle$, $|0110\rangle$, $|0111\rangle$, $|1000\rangle$, $|1001\rangle$, $|1010\rangle$, $|1011\rangle$, $|1100\rangle$, and $|1101\rangle$\}, the partial transpose of the density matrix with respect to subsystem $D$ exhibits a block-matrix structure, which can be written as
\begin{eqnarray}\label{w26}
\scalebox{0.8}{$
\rho^{T_{D}}_{ABCD}=
\left(\!\!\begin{array}{cccccccccccccccc}
Z_{1}&0&Z_{1}&0&0&Z_{0}&Z_{1}&0&0&Z_{0}&0&Z_{0}\\
0&Z_{0}&0&Z_{0}&0&0&0&Z_{0}&0&0&0&0\\
Z_{1}&0&Z_{1}&0&0&Z_{0}&Z_{1}&0&0&Z_{0}&0&Z_{0}0\\
0&Z_{0}&0&Z_{0}&0&0&0&Z_{0}&0&0&0&0\\
0&0&0&0&Z_{0}&0&0&0&Z_{0}&0&0&Z_{0}\\
Z_{0}&0&Z_{0}&0&0&Z_{2}&Z_{0}&0&0&Z_{2}&0&Z_{2}\\
Z_{1}&0&Z_{1}&0&0&Z_{0}&Z_{1}&0&0&Z_{0}&0&Z_{0}&\\
0&Z_{0}&0&Z_{0}&0&0&0&Z_{0}&0&0&0&0\\
0&0&0&0&Z_{0}&0&0&0&Z_{0}&0&Z_{0}&0\\
Z_{0}&0&Z_{0}&0&0&Z_{2}&Z_{0}&0&0&Z_{2}&0&Z_{2}\\
0&0&0&0&Z_{0}&0&0&0&Z_{0}&0&Z_{0}&0\\
Z_{0}&0&Z_{0}&0&0&Z_{2}&Z_{0}&0&0&Z_{2}&0&Z_{2}\\
\end{array}\!\!\right)
$}.
\end{eqnarray}
Due to the complexity of the eigenvalue spectra of the partially transposed density matrices, the analytical expressions of the negativities $N_{A(BCD)}$ and $N_{D(ABC)}$ are not presented explicitly.

\begin{figure}
\begin{minipage}[t]{0.5\linewidth}
\centering
\includegraphics[width=3.0in,height=5.2cm]{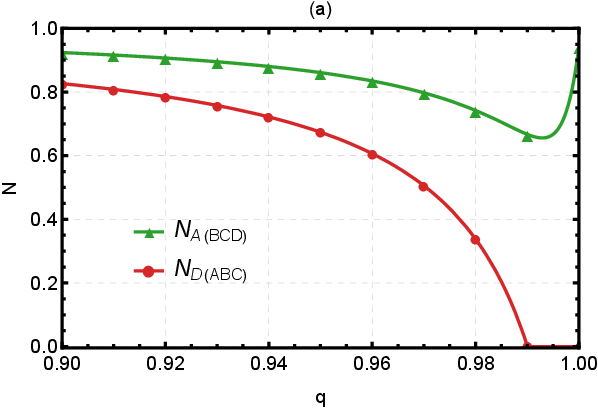}
\label{fig2a}
\end{minipage}%
\begin{minipage}[t]{0.5\linewidth}
\centering
\includegraphics[width=3.0in,height=5.2cm]{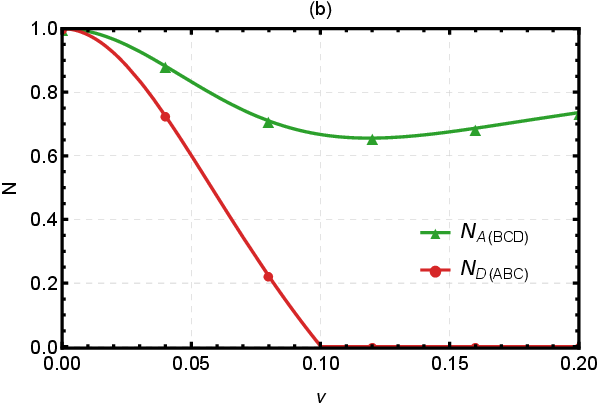}
\label{fig2b}
\end{minipage}%

\caption{Dependence of the $1-3$ bipartite entanglement $N_{A(BCD)}$ and $N_{D(ABC)}$ on the acceleration parameter $q$ and the effective coupling $\nu$. Panel (a) shows their variation as functions of $q$ for fixed $\nu^{2}=0.01$, while panel (b) displays their dependence on $\nu$ for fixed $q=0.99$.}
\label{Fig2}
\end{figure}

In Fig.\ref{Fig2}, we plot the negativities of the four-qubit Dicke state as functions of the acceleration parameter $q$ and the effective coupling $\nu$. As shown in Fig.\ref{Fig2}(a), the negativity $N_{A(BCD)}$, which characterizes the entanglement between Alice and the remaining subsystem, decreases rapidly with increasing $q$ at low accelerations, but does not vanish completely. Instead, it gradually increases again as $q\rightarrow1$, approaching a finite nonzero value in the infinite-acceleration limit,  meaning  that the Unruh effect can enhance $1-3$ bipartite entanglement within a finite parameter regime. By contrast, the negativity associated with the accelerated observer, $N_{D(ABC)}$, exhibits a markedly different behavior. Although it remains close to unity at small accelerations, it decreases rapidly and eventually undergoes entanglement sudden death  with increasing $q$, showing that the entanglement involving non-inertial observers is highly sensitive to the thermal noise generated by the Unruh effect. This asymmetry reflects the strong observer dependence of quantum correlations in relativistic quantum information theory. Fig.\ref{Fig2}(b) shows the dependence of the negativities on the effective coupling $\nu$ for fixed $q=0.99$. In this case, $N_{A(BCD)}$ initially decreases with increasing $\nu$ and subsequently exhibits a slight enhancement, whereas $N_{D(ABC)}$ decays monotonically and vanishes at sufficiently large coupling. These results reveal the dual role of the Unruh effect in Dicke-state entanglement, namely, that relativistic motion can both suppress and enhance bipartite correlation depending on the subsystem partition and parameter regime.

\subsection{Whole entanglement measures in relativistic motion}

To characterize the global multipartite entanglement of the four-qubit system, we employ the residual entanglement measure known as the $\pi_{4}$-tangle, which is constructed from the monogamy relation of negativity. The residual tangles associated with each subsystem are defined as
\begin{align}
\pi_{A} &= N^{2}_{A(BCD)} - N_{AB}^{2} - N_{AC}^{2} - N_{AD}^{2}, \label{eq:17} \\
\pi_{B} &= N^{2}_{B(ACD)} - N_{BA}^{2} - N_{BC}^{2} - N_{BD}^{2}, \label{eq:18} \\
\pi_{C} &= N^{2}_{C(ABD)} - N_{CA}^{2} - N_{CB}^{2} - N_{CD}^{2}, \label{eq:19} \\
\pi_{D} &= N^{2}_{D(ABC)} - N_{DA}^{2} - N_{DB}^{2} - N_{DC}^{2}, \label{eq:20}
\end{align}
from which the total four-partite entanglement measure is obtained as
\begin{equation}
\pi_{4} = \frac{1}{4}\left(\pi_{A} + \pi_{B} + \pi_{C} + \pi_{D}\right). \label{eq:21}
\end{equation}
The $\pi_{4}$-tangle provides a symmetric quantifier of intrinsic  multipartite entanglement by combining the residual contributions associated with all subsystem partitions, thereby characterizing the overall distribution of four-qubit quantum correlations in the relativistic system.

\begin{figure}
\begin{minipage}[t]{0.5\linewidth}
\centering
\includegraphics[width=3.0in,height=5.2cm]{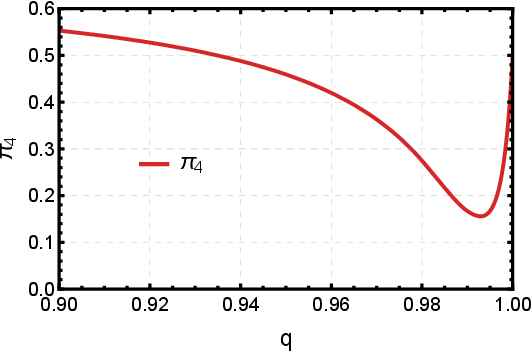}
\label{fig3}
\end{minipage}%
\caption{The $\pi_4$-tangle  as functions of the acceleration parameter $q$ for fixed effective coupling $\nu^{2}=0.01$.}
\label{fig3}
\end{figure}

Fig.\ref{fig3} shows the behavior of the global multipartite entanglement measure $\pi_{4}$ as  functions of the acceleration parameter $q$ for fixed $\nu^{2}=0.01$. We find that the $\pi_{4}$-tangle exhibits a distinct non-monotonic evolution under relativistic motion: it initially decreases with increasing acceleration, reaches a minimum, and subsequently increases toward a finite nonzero asymptotic value. This behavior is qualitatively different from that previously observed for GHZ and W states \cite{UD20}, whose multipartite entanglement decreases monotonically with acceleration and eventually undergoes entanglement sudden death under the Unruh effect. The distinct behavior of the Dicke state originates from its intrinsic entanglement structure. Unlike GHZ states, whose multipartite entanglement is concentrated in a single global coherence, or W states, whose correlations are predominantly distributed in pairwise sectors, the Dicke state possesses a more balanced and symmetrically distributed multipartite entanglement. Such a distributed correlation structure enables acceleration-induced mode mixing to redistribute quantum correlations more efficiently among different multipartite sectors. Consequently, rather than being completely destroyed by the Unruh thermal noise, part of the multipartite correlations is transferred back into the physically accessible sector, leading to the partial recovery of the $\pi_{4}$-tangle at large accelerations. Therefore, the Unruh effect does not merely degrade multipartite entanglement, but can also partially enhance it within certain parameter regimes.
The finite residual multipartite entanglement in the infinite-acceleration limit arises because the Unruh effect redistributes, rather than completely destroys, the multipartite quantum correlations of the Dicke state. Consequently, a finite fraction of the correlations remains encoded in the physically accessible sector even under extremely strong acceleration, ensuring the survival of residual entanglement. The persistence of multipartite entanglement in the Dicke state therefore demonstrates its substantially stronger robustness against Unruh-induced decoherence and suggests that Dicke states constitute particularly promising multipartite quantum resources for relativistic quantum information processing and quantum communication in non-inertial settings. Although observing the genuine Unruh effect remains experimentally challenging, the present detector-based scenario can, in principle, be investigated using quantum simulation platforms, where effective Unruh-like phenomena have already been proposed or realized \cite{ZZ28}.

The underlying mechanism can be understood as the competition between acceleration-induced thermal decoherence and the redistribution of multipartite quantum correlations caused by mode mixing. At relatively small accelerations, the Unruh thermal noise dominates, leading to a rapid loss of the initially shared entanglement. As the acceleration further increases, however, the Bogoliubov mode mixing associated with the accelerated detector redistributes quantum correlations among different multipartite sectors rather than simply destroying them. Consequently, part of the quantum correlations not captured in the low-acceleration regime are transferred back into the physically accessible multipartite entanglement, giving rise to the observed partial recovery of the $\pi_4$-tangle at large accelerations. Therefore, the non-monotonic behavior originates from the interplay between Unruh-induced decoherence and correlation redistribution, rather than from a genuine creation of new entanglement.

\begin{figure}
\begin{minipage}[t]{0.5\linewidth}
\centering
\includegraphics[width=3.0in,height=6.24cm]{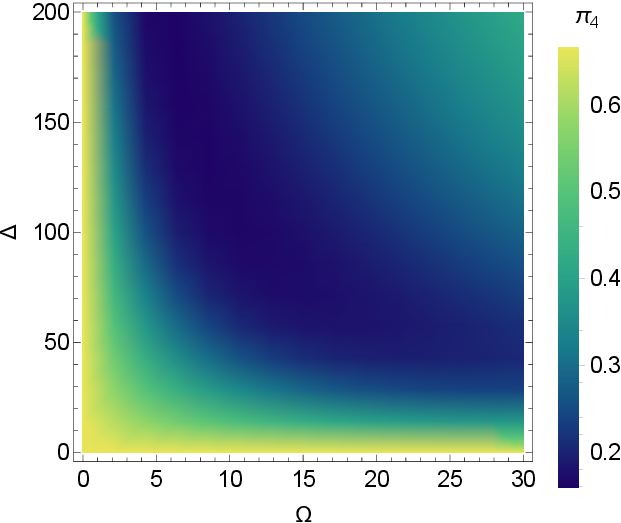}
\label{fig4}
\end{minipage}%
\caption{The $\pi_{4}$-tangle of the four-qubit Dicke state as functions of the interaction time $\Delta$ and the energy gap $\Omega$, with $\varepsilon^{2} = 8\pi^{2} \times 10^{-6}$, $\kappa = 0.02$, and $q = 0.99$.}
\label{Fig4}
\end{figure}

Fig.\ref{Fig4} shows the dependence of the global multipartite entanglement measure $\pi_{4}$ on the detector energy gap $\Omega$ and the interaction time $\Delta$. The results indicate that the $\pi_{4}$-tangle exhibits a clear non-monotonic behavior with respect to both parameters, characterized by an initial decrease followed by a subsequent increase. This demonstrates that the detector-field interaction plays a crucial role in regulating the redistribution of multipartite quantum correlations in relativistic motion. In particular, the competition between acceleration-induced decoherence and detector-field coupling effects leads to a partial recovery of residual multipartite entanglement in certain parameter regimes.  Moreover, the strong dependence of the entanglement dynamics on the detector parameters $\Omega$ and $\Delta$ indicates that both the detector configuration and subsystem structure play crucial roles in relativistic quantum information tasks. By exploiting the distinctive robustness of the global multipartite entanglement in Dicke states, it may be possible to design experimentally accessible advanced detection schemes based on artificial two-level atoms with tunable energy gaps, thereby preserving multipartite quantum correlations even in the presence of Unruh-induced thermal noise. These results suggest that Dicke states constitute promising multipartite quantum resources for implementing robust quantum communication and quantum information processing protocols in non-inertial and relativistic quantum field settings.

\section{Conclusions}
In this work, we have investigated the dynamics of  global tetrapartite entanglement in a four-qubit Dicke state within the framework of the Unruh--DeWitt detector model, where one detector undergoes uniform acceleration while the remaining detectors remain inertial. Motivated by the question of whether multipartite entanglement in Dicke states necessarily follows the conventional monotonic degradation behavior under relativistic effects, we analyzed the evolution of  the global multipartite entanglement quantified by the $\pi_{4}$-tangle. Our results reveal that the multipartite entanglement dynamics of the Dicke state differ fundamentally from those of previously studied multipartite quantum resources \cite{SDF27,SDF28,SDF31,ZZ4,ZZ5,RM1,RM2,RM3,RM4,RM5,RM6,UD20}.
In particular, the multipartite entanglement exhibits a pronounced non-monotonic behavior: with increasing acceleration, it initially decreases, reaches a minimum, and subsequently increases toward a finite asymptotic value. This demonstrates that the Unruh effect does not merely degrade multipartite quantum correlations, but can also partially enhance them within certain parameter regimes. Therefore, our results provide a refined understanding of the influence of relativistic motion on multipartite entanglement and show that the conventional picture of monotonic multipartite entanglement degradation is not universally valid.

Our second motivation was to explore whether Dicke states can serve as more robust multipartite quantum resources against Unruh-induced decoherence. The results clearly support this conclusion. Unlike GHZ and W states, whose multipartite entanglement decreases monotonically and eventually suffers entanglement sudden death under relativistic effects \cite{UD20}, the Dicke state exhibits a distinct non-monotonic evolution and retains a finite amount of residual multipartite entanglement even in the large-acceleration limit. This persistence originates from acceleration-induced mode mixing and the redistribution of quantum correlations, which enable part of the multipartite entanglement to survive despite the presence of Unruh thermal noise. Furthermore, we showed that the entanglement dynamics depend sensitively on the detector energy gap and interaction time, indicating that detector configurations can effectively regulate the redistribution and robustness of multipartite quantum correlations. These features suggest that Dicke states possess substantially stronger robustness against relativistic decoherence and therefore constitute particularly promising multipartite quantum resources for relativistic quantum communication and quantum information processing tasks. In particular, the robustness of Dicke-state global multipartite entanglement may provide a realistic route toward experimentally accessible relativistic quantum technologies based on artificial two-level atoms with tunable energy gaps, capable of preserving quantum correlations even in non-inertial and relativistic quantum field environments.

\begin{acknowledgments}
This work is supported by the National Natural Science Foundation of China (12575056),  the Natural Science Foundation of Hunan Province under Grant No.2025JJ60019, the Natural
Science Foundation of the Liaoning Scientific Committee No. 2026-MS-281, and LiaoNing Revitalization Talents Program (XLYC2503099).	
\end{acknowledgments}


\end{document}